\documentclass[12pt,manuscript]{aastex}

\usepackage{natbib,aas_macros}
\usepackage{amsmath}

\newcommand{\ep}{\epsilon}

\usepackage{parskip}

\shorttitle{Accretion onto planets}
\shortauthors{Uribe, Klahr and Henning}

\begin{document}

\title{Accretion of gas onto gap-opening planets and circumplanetary flow structure in magnetized turbulent disks}

\author{A. L. Uribe\footnote{Currently at University of Chicago}, H. Klahr, and Th. Henning}
\affil{Max-Planck-Institut f\"{u}r Astronomie, Heidelberg, Germany.}
\email{uribe@oddjob.uchicago.edu}

\begin{abstract}
We have performed three-dimensional magneto-hydrodynamical simulations of stellar accretion disks, using the PLUTO code, and studied the accretion of gas onto a Jupiter-mass planet and the structure of the circumplanetary gas flow after opening a gap in the disk. We compare our results with simulations of laminar, yet viscous disks with different levels of an $\alpha$-type viscosity. In all cases, we find that the accretion flow across the surface of the Hill sphere of the planet is not spherically or azimuthally symmetric, and is predominantly restricted to the mid-plane region of the disk. Even in the turbulent case, we find no significant vertical flow of mass into the Hill sphere. The outer parts of the circumplanetary disk are shown to rotate significantly below Keplerian speed, independent of viscosity, while the circumplanetary disk density (therefore the angular momentum) increases with viscosity. For a simulation of a magnetized turbulent disk, where the global averaged alpha stress is $\alpha_{MHD}=10^{-3}$, we find the accretion rate onto the planet to be $\dot{M}\approx 2\times10^{-6}M_{{J}}yr^{-1}$ for a gap surface density of $12 g  cm^{-2}$. This is about a third of the accretion rate obtained in a laminar viscous simulation with equivalent $\alpha$ parameter.
\end{abstract}


\section{INTRODUCTION}\label{intro}

Gas giants can form as a result of core formation by planetesimal accretion followed by formation of the gas envelope through accretion from the circumstellar disk \citep{1980PThPh..64..544M}. \citet{1996Icar..124...62P} distinguished three phases of this process, using numerical simulations of core accretion and envelope evolution. A first phase marked by the fast accretion of solids onto a core until the feeding zone of the planet is mostly evacuated \citep{1982P&SS...30..755S}; a second phase where gas and solid accretion is low and constant; a third stage when the core mass equals the envelope mass, leading to the contraction of the envelope and the onset of runaway gas accretion \citep{1980PThPh..64..544M}. Migration of the planet might allow for an extension of the feeding zone, while gap formation might lead to a mass limit for gap opening planets \citep{2005A&A...434..343A}. In the outer parts of the disk, giant planets could form as a result of the collapse of a gravitationally unstable disk clump. This mechanism requires a very massive disk that can cool effectively on timescales of a few local orbital periods \citep{1997Sci...276.1836B,2002Sci...298.1756M,2005ApJ...621L..69R,2007prpl.conf..607D}. 
\par
Monte Carlo models of planet formation produce synthetic populations of extrasolar planets, with a large diversity of initial conditions. These models have been successful in reproducing key features of the observed distribution of exoplanets (e.g. \citet{2008PhST..130a4022B,2009A&A...501.1161M,2009A&A...501.1139M,2008IAUS..249..223I}). The calculations usually include one-dimensional disk evolution and core/envelope structure models. The accretion of planetesimals and gas onto an already formed proto-core is included, using a given prescription for the accretion rates of gas and rocky materials onto the planet \citep{2004A&A...417L..25A}. For this reason, an accurate estimation and parameterization of the accretion rates of gas onto planets for a variety of conditions is necessary to correctly calculate the formation time and the limiting mass of the giant planetary population.  

So far, the accretion of gas onto planets has been modeled using two different approaches. On one side, one dimensional models have been used to calculate the radial structure of the envelope and the accretion onto a rocky core. These models can include effects such as the dust opacity of the envelope, the release of energy of infalling planetesimals into the envelope and the thermal feedback of the planet \citep{2000ApJ...537.1013I,2005Icar..179..415H}. These models can include the disk evolution only in a restricted way, and assume a certain model of the outer envelope (as a boundary condition) that is spherically symmetric.
\par
On the other hand, two/three-dimensional simulations of accretion disk with accreting planets aim to estimate the structure of the flow around the planet and how much mass the disk is capable of feeding to the planet. However, most of these simulations do not include the radiative feedback from the planet and a detailed model of the inner envelope. \citet{2002ApJ...580..506T} and \citet{2003ASPC..294..323D,2003ApJ...586..540D} used high-resolution two-dimensional simulations to study the detailed flow pattern around the planet, and the circumplanetary disk. They showed that inside the planet's Roche lobe, accretion in the circumplanetary disk is powered mainly by energy dissipation of circulating matter at the spiral shock. Outside the Roche lobe, gas flows onto the planet through "accretion bands" located between the horseshoe flow and the passing-by flow (although the detailed structure depends strongly on the sound speed). The accretion timescale 
\begin{equation}
\tau_{acc} = \frac{M_{p}}{\dot{M_{p}}},
\end{equation} 
has been measured to be around $10^{4}-10^{5}$ yr, and the accretion rate of a Jupiter-mass planet has been found to be of the order of $10^{-5} M_{J}/yr$ on a disk with $M_{d}=0.01M_{\odot}$ \citep{1999ApJ...514..344B,1999ApJ...526.1001L,2001ApJ...547..457K,2003MNRAS.341..213B}. Three-dimensional simulations including radiation transfer have found similar accretion rates and have shown the formation of a thick ($H/r\approx0.5$) circumplanetary disk \citep{2006A&A...445..747K}. 
\par
The properties of the circumplanetary disk have been the subject of various numerical and semi-analytical studies. \citet{2009MNRAS.397..657A} used radiation-hydrodynamics numerical simulations to study the planet-disk system, and derived important properties such as the power law dependence of density with radius, and the scale height of the stellar disk. Their calculations show that the disk density falls with radius as $r^{-2}$ to $r^{-3/2}$, and the disk is fairly thick, with $H/r>0.2$, with no significant flaring. \citet{2010AJ....140.1168W} developed a model for the circumplanetary disk that includes the contraction of the planet envelope and the properties of the inflowing gas material from the disk. They find various phases of gas accretion and collapse where the circumplanetary disk can either provide mass to the planet or act as a "spin-out disk" where the mass flux is outwards. The inner circumplanetary disk rotates Keplerian up to a small fraction of the Hill sphere and then rotation becomes sub-Keplerian in the outer region. The issue of the circumplanetary disk size, or the truncation mechanism of the circumplanetary disk, is also crucial for the fate of the solid material. The inner Keplerian rotating thick disk is truncated at a fraction of the Hill sphere ($\propto 0.4r_{h}$), but this seems to depend on the viscous mechanism acting in the circumplanetary disk \citep{2011MNRAS.413.1447M,2003MNRAS.341..213B}.
\par
Recent work by \citet{2007ApJ...667..557T} suggests a different picture for the accretion flow in an inviscid isothermal circumplanetary disk. In this case, the bulk of the accretion does not occur in the disk mid-plane region, but instead at high latitudes in the planet Hill sphere. In the equatorial region, an outflow is present through the Lagrangian points L1 and L2. This flow structure is similar to that observed by \citet{2006A&A...445..747K} in radiative non-viscous simulations. 

\par
In this paper we perform numerical simulations to study the accretion rate of gas onto giant gap-opening planets and the circumplanetary disk in turbulent magnetized disks using, for the first time, three-dimensional global stellar disk simulations. The turbulence in the disk is generated by the magneto-rotational instability (MRI) \citep{1991ApJ...376..214B}. A minimum degree of ionization leading to a good coupling between the gas and the magnetic field in the stellar disk is required for the MRI to function. In the cold and dusty mid-plane regions the ionization will be low, possibly producing a dead-zone, while on the outer parts of the disk, cosmic rays can penetrate allowing for the necessary ionization degree \citep{2010ApJ...708..188T,2010A&A...515A..70D,2011ApJ...735..122F}. A planet gap, however, might lower the local density to a sufficient level to allow for ionization of the mid-plane regions, leading to an MRI active zone. Therefore, this type of turbulence could conceivable be still present in the planet corotation region.
\par
 We compare the accretion rate onto a planet in an MRI-turbulent disk with that in a viscous laminar disk and we examine the accretion structure and mass inflow into the Hill sphere, where material passes through the circumplanetary disk to be accreted by the planet. The viscosity provided by the turbulence in the stellar disk (either modeled by an $\alpha$ prescription or by "real" magnetic turbulence) is acting indirectly in the circumplanetary disk. There is no extra viscosity applied to the circumplanetary disk, nor is this disk MRI unstable on its own. All the effects we observe are due to the viscosity of the circumstellar disk, be this a laminar kinematic viscosity, or a magnetic turbulent transport. Note that the MRI is not resolved in the circumplanetary disk, and the question of what mechanism drives accretion in this disk is still unresolved. The effective transport comes from the transport of angular momentum in the stellar disk. We also concentrate strictly on the phase when a gap has been cleared in the disk, and study the efficiency of the stellar disk in providing mass to the forming protoplanet. The choice of simulating a locally isothermal disk is justified for simulations with a Jupiter mass planet since at this planet mass the accretion rate approaches the isothermal result and is independent of the opacity treatment \citep{2009MNRAS.397..657A}. Additionally, for this mass range, the accretion rate onto the planet depends on the ability of the circumstellar disk to provide material, which is the effect we study in this paper.
\par
The paper is organized as follows. In Section 2, we describe the computational setup, boundary and initial conditions, and the parameters we use in our simulations. We also describe the prescription for the mass accretion onto the planet. In Section 3.1, we present our results on the three-dimensional structure of the accretion flow into the Hill sphere. Section 3.2 contains the results on the mass accretion rates for the different simulations, while Section 3.3 shows additional details of the accretion flow around the planet. Finally, we discuss and summarize our results in Sections 4 and 5.

\section{COMPUTATIONAL SETUP}\label{setup}

Simulations were performed using the finite volume fluid dynamics code PLUTO \citep{2007ApJS..170..228M}.  In the code, time stepping is done using a second order Runge Kutta scheme, while the spatial integration is performed using linear interpolation through the second order TVD scheme. The Riemann fluxes are computed using the HLLC and HLLD solvers for the HD and MHD cases, respectively. The code uses the Constrained Transport method for preserving a divergence-free magnetic field \citep{2005JCoPh.205..509G}. The numerical setup for the MHD case follows the setup presented in
\citet{2010A&A...516A..26F}. 
\par
We use spherical coordinates $(r,\theta,\phi)$ and the domain is given by $r\in [1,10]$(in units of AU), $\theta \in [\pi/2 -0.3,\pi/2 +0.3]$ and $\phi \in [0,2\pi]$. The grid resolution is $(N_{r},N_{\theta},N_{\phi})=(256,128,256)$ and it is centered in the center of mass of the planet-star system. 
\par
The circumstellar gas disk is initially in sub-Keplerian rotation around a solar mass star. The azimuthal velocity is given by
\begin{equation}
v_{\phi}= \sqrt{v_{k}^{2} - c_{s}^{2}(a - 2b)},
\end{equation}
where $v_{k}$ is the Keplerian velocity and $a=3/2$ and $b=0.5$ are the exponents of the radial power law distribution of the density $\rho\propto r^{-a}$ and sound speed $c_{s}=c_{0}(r\sin\theta)^{-b}$. The initial density distribution is given by
\begin{equation}
\rho(r,\theta) = (r\sin\theta)^{-3/2}\exp\left(\frac{\sin\theta-1}{c_{0}^{2}}\right).
\label{eq:dens_prof2}
\end{equation}
The disk is described by a locally isothermal equation of state $P=c_{s}^{2}\rho$. The ratio of the pressure scale height $h$ to the radial coordinate of the disk $h=c_{s}/v_{k}$ is taken to be a constant such that $h=H/(r\sin\theta)=0.07$.\\
The gravitational potential of the planet is given by a softened point-mass potential
\begin{equation}
\Phi_{p}(\mathbf{r}) = -\frac{GM}{(|\mathbf{r}-\mathbf{r}_{p}|^2 + \ep^{2})^{1/2}}
\end{equation}
where $\ep$ is the softening parameter, needed to avoid numerical divergence near the position of the planet. For all the simulations $\ep$ is set to be a fraction of the Hill radius $\ep=kr_{p}(M_{p}/3)^{1/3}$ with $k=0.3$. Distances are given in units of $r_{0}=1AU$, density is given in units of $\rho_{0}=1\times10^{-12} g cm^{-3}$, and velocity is given in units of Keplerian speed at $1AU$,  $v_{0}=v_{k}(1AU)$. The density has been scaled such that the total disk mass is $0.01M_{star}$. Magnetic fields are given in units of $B_{0}=\sqrt{4\pi \rho_{0} v^{2}_{0}} $. The equations of motion of the planet are solved at each timestep with a leap frog integrator.

\subsection{Boundary conditions}
The boundary conditions for the velocities and magnetic field are periodic in the vertical ($\theta$ boundary) and azimuthal directions and reflective in the radial direction, except for the transverse magnetic field component, which reverses its sign at the radial boundary. Buffer zones are defined at the radial boundaries to avoid boundary effects, where for $1<r<2$ the magnetic resistivity is given by $\eta=2\times10^{-4}(2-r)$ and for $9<r<10$ the resistivity is $\eta=1\times10^{-4}(r-9)$. 

\subsection{Initial conditions, gap opening and viscosity}

\begin{figure*}[h!]\centering
  \includegraphics[width=15cm]{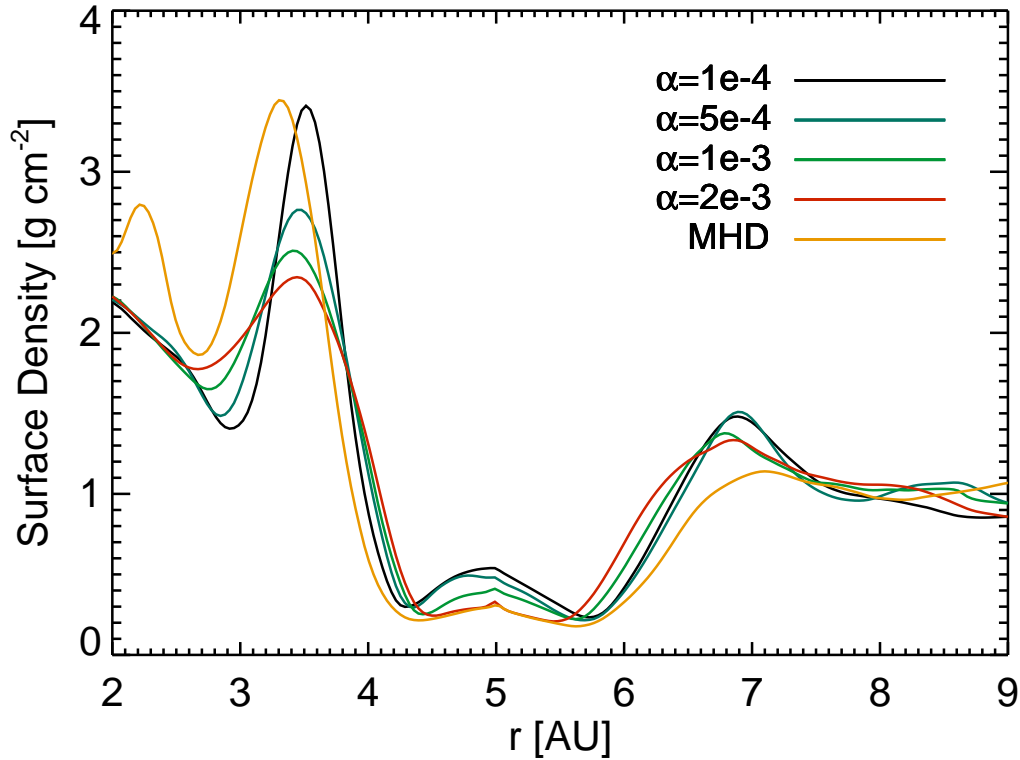}
   \caption{Initial conditions before the planet starts accreting for the laminar viscous disk simulations and the MHD simulation. The gap in the MHD simulation is found to be wider as compared to all the viscous simulations. The figure shows the density after 100 orbital periods have elapsed. }\label{fig:gap}
\end{figure*}

Since our aim is to find the accretion rate onto the planet only in the stage when the gap is cleared, the planet is allowed to accrete gas after 100 orbital periods at $5AU$ have elapsed (see Figure \ref{fig:gap}). At this stage, a gap has been cleared, and the density has been reduced by more than 95\%. For the hydrodynamical simulations, viscosity is added explicitly as a source term in the momentum equation. We use an $\alpha$-type kinematic viscosity in the circumstellar disk which is given by $\nu=\alpha c_{s}H$, where $\alpha$ takes values of $2\times10^{-3}, 1\times10^{-3},  5\times10^{-4}$, and $1\times10^{-4}$. 

\subsection{Accretion prescription}
 The accretion onto the planet is modeled by removing a fraction of the mass inside a portion of the Hill sphere at each time step. At each timestep the new density $\rho^{'}$ is given by
\begin{equation}
\rho^{'}(\mathbf{r}) = \left(1 - \frac{\Delta t}{t_{a}}\right)\rho(\mathbf{r}).
\end{equation}
The accreted mass by the planet in timestep $\Delta t$ is $\Delta M=(\rho(\mathbf{r})\Delta t t_{a}^{-1}) r^{2}sin(\theta)dr d\theta d\phi$. The planet accretion rate for timestep $\Delta t$ is calculated as the accreted mass divided by the timestep $\Delta M/\Delta t$. The factor $t_{a}$ represents the accretion timescale in which the Hill sphere is emptied if there was no gas flowing in from the disk. This is chosen to be $t_{a}=1\Omega_{1AU}^{-1}$ inside the inner half of the Hill sphere and $t_{a}=2\Omega_{1AU}^{-1}$ in the outer half of the Hill sphere (where $\Omega_{1AU}^{-1}$ is the Keplerian angular frequency at $1AU$). A density floor is applied to the simulations with magnetic fields, where the density is not allowed to drop below $10^{-19}g cm^{-3}$. Nevertheless, the density in the simulation never reaches this value. The magnetic field is not modified as the density is reduced inside the Hill sphere in order to preserve a divergence-free field. The accretion rate has been shown to be dependent on the accretion radius (the distance from the planet up to which mass is removed) and on the accretion timescale parameter $t_{a}$. \citet{2002ApJ...580..506T} showed that the accretion radius should be small ($\approx0.1r_{h}$) and the accretion timescale should be on the order of the orbital period, in order to obtain converged results. Because of our lower resolution, we take most of the mass from within the inner half of the Hill sphere (our resolution element is $0.035r_{0}$ or $0.1r_{hill}$ ). This prescription has also been used in previous studies of gas accretion and migration by giant planets \citep{2001ApJ...547..457K}. We also checked that our results remain valid if one restricts the accretion radius down to $0.4r_{h}$. We have also verified that we obtain the same results if we extend the accretion timescale to $t_{a}=10$.

\section{RESULTS}\label{res}

\subsection{Structure of the envelope and mass inflow}

\begin{figure*}[ht]
\begin{minipage}[b]{.5\linewidth}
  \includegraphics[width=\linewidth]{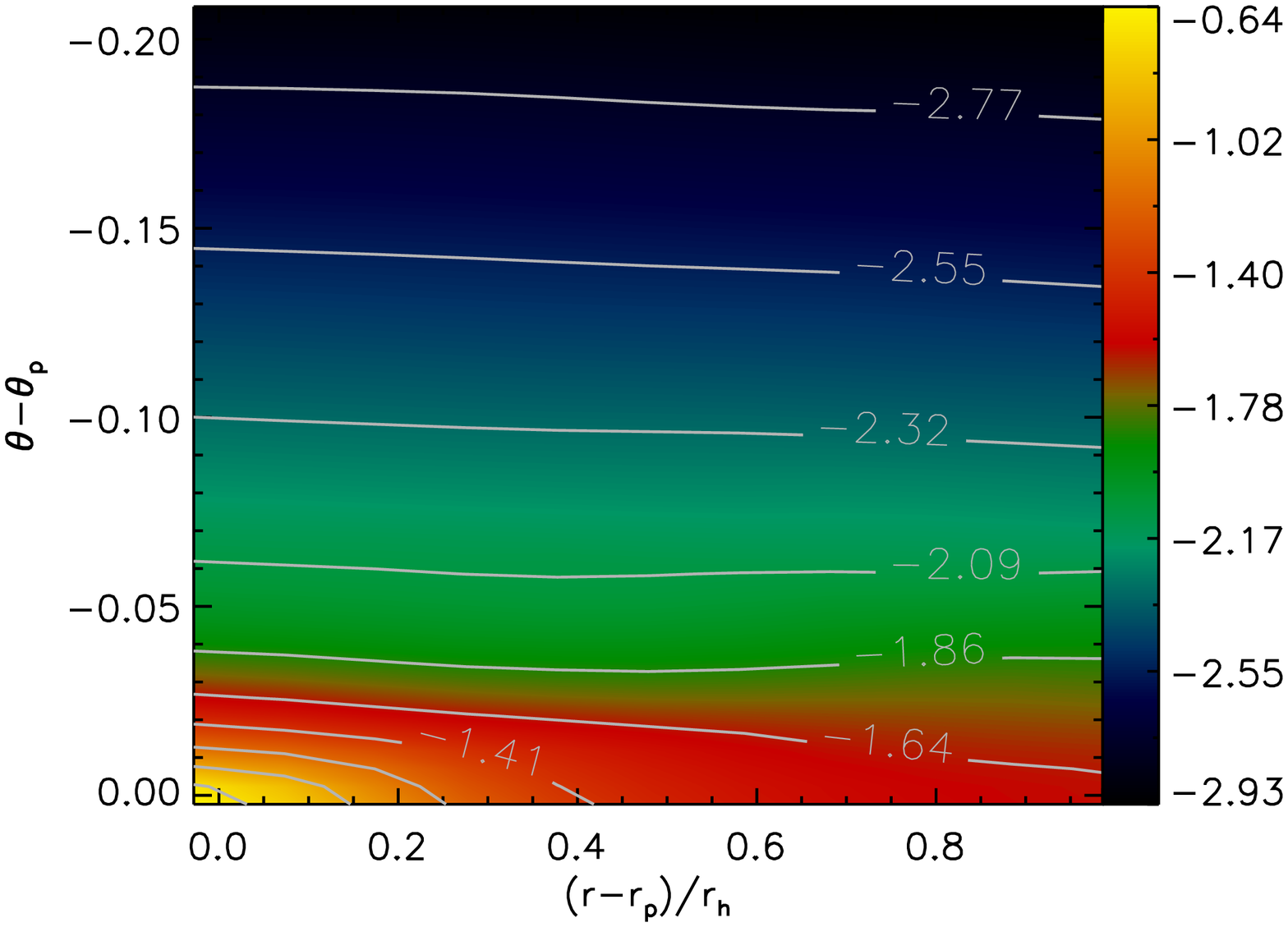}
 \end{minipage}
\begin{minipage}[b]{.5\linewidth}
  \includegraphics[width=\linewidth]{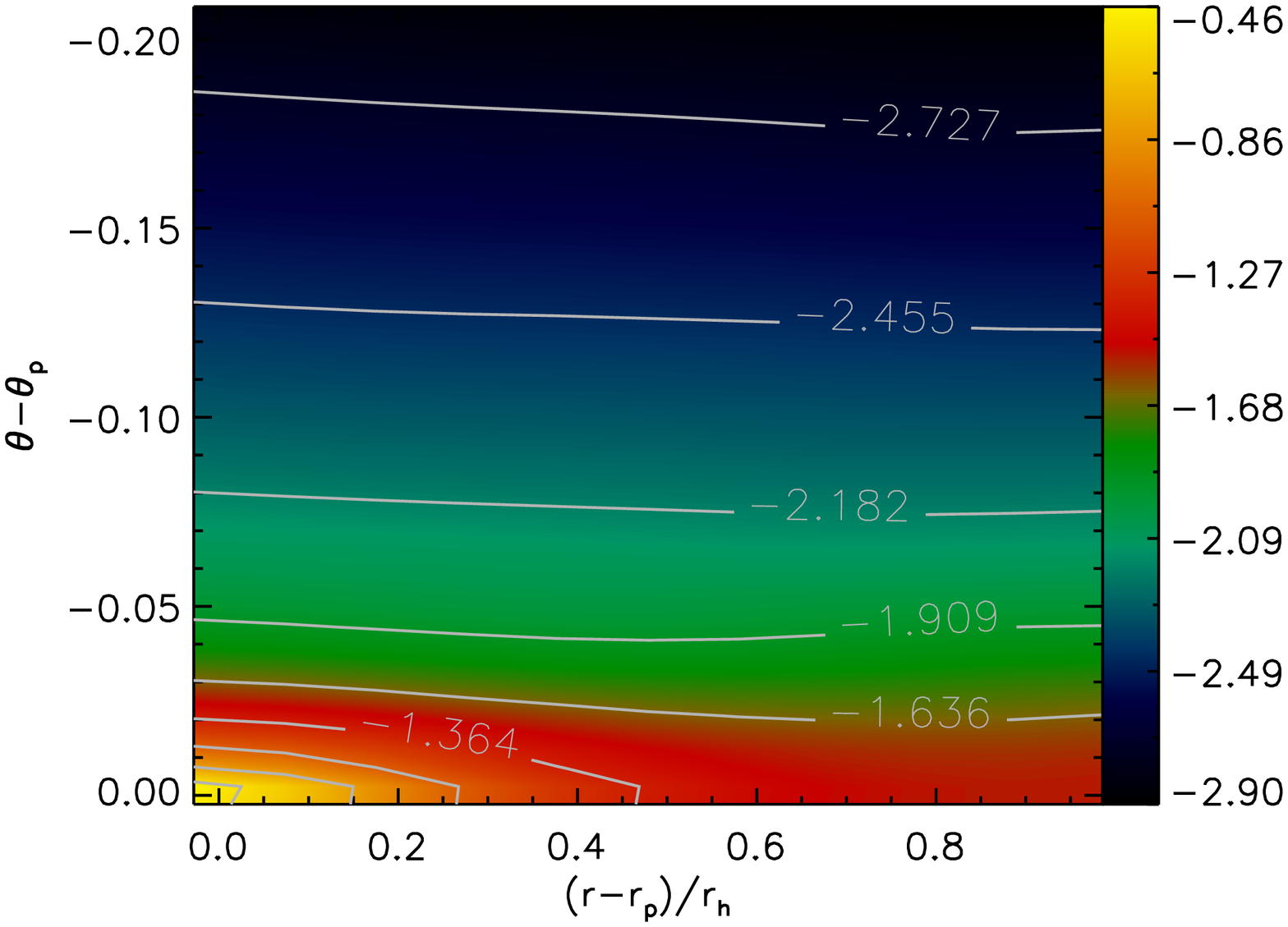}
 \end{minipage}
\begin{minipage}[b]{.5\linewidth}
  \includegraphics[width=\linewidth]{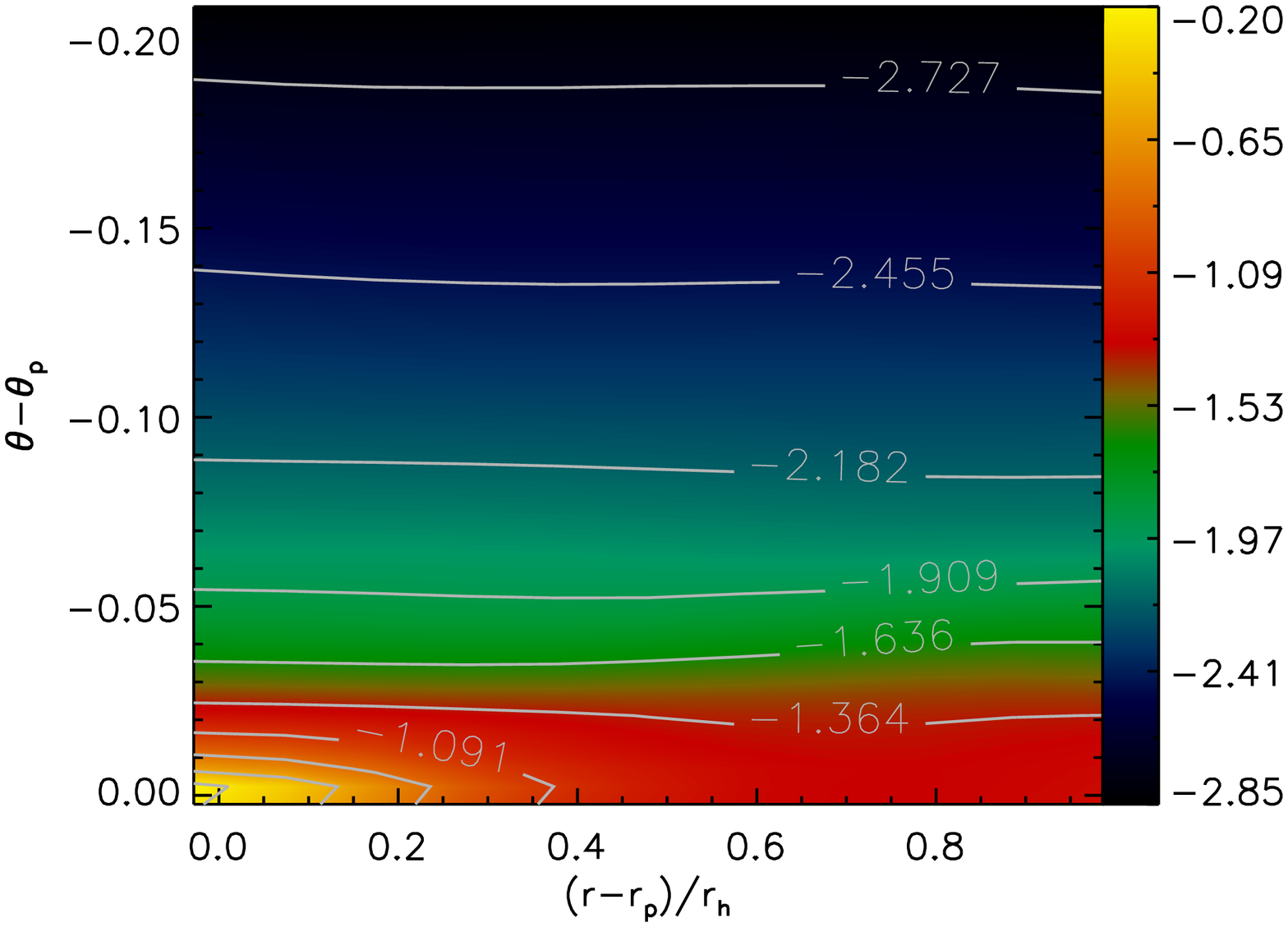}
 \end{minipage}
\begin{minipage}[b]{.5\linewidth}
  \includegraphics[width=\linewidth]{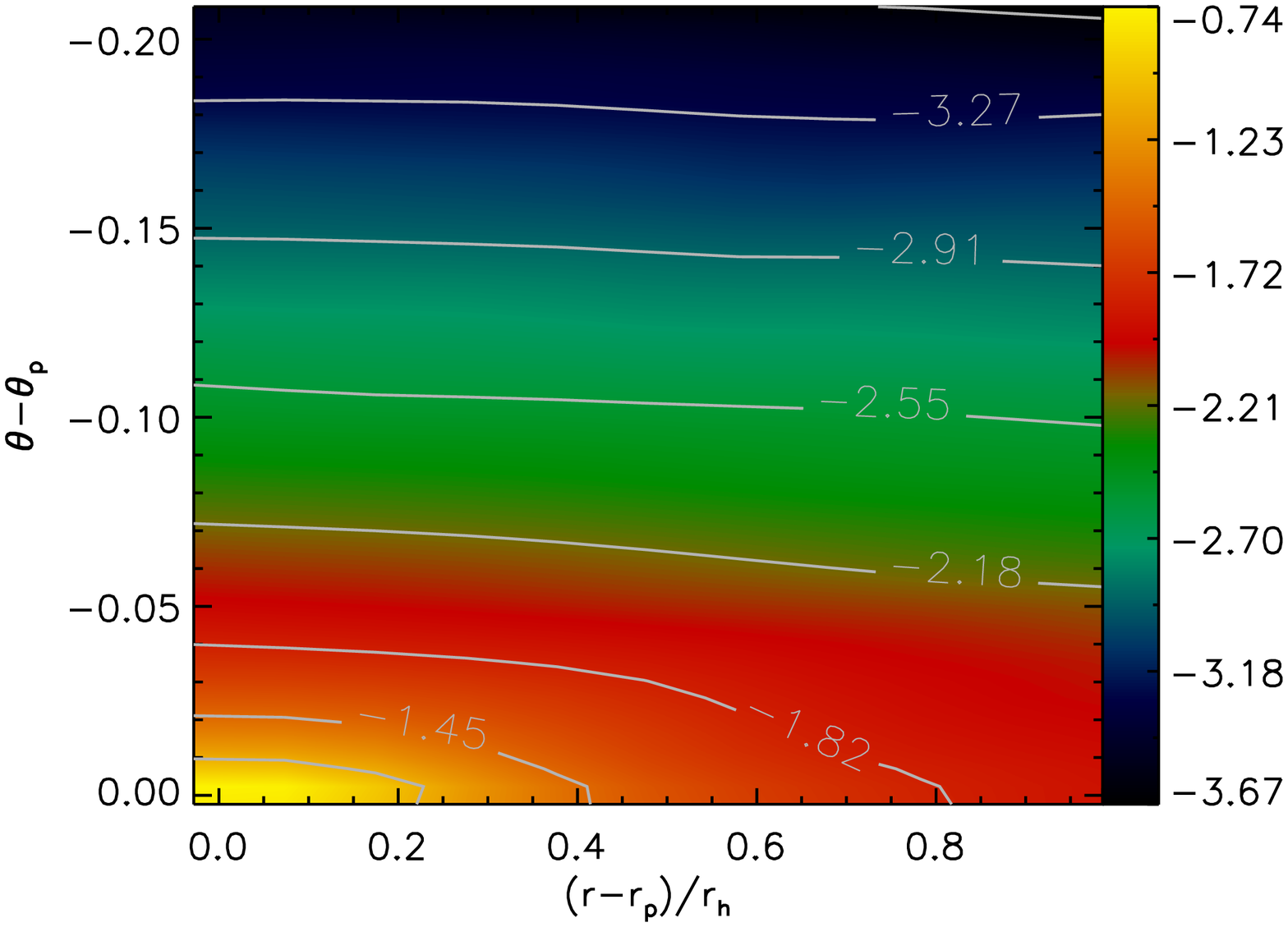}
 \end{minipage}
\caption{Density structure of the disk around the planet (in the radial and vertical direction). The planet is located at the origin of the coordinate system. Color and contour lines show the logarithm of the density. Top left: $\alpha=5\times10^{-4}$. Top right: $\alpha=10^{-3}$. Bottom left: $\alpha=2\times10^{-3}$. Bottom right: MHD simulation.}\label{fig:f1}
\end{figure*}

\begin{figure*}[h!]\centering
  \includegraphics[width=\linewidth]{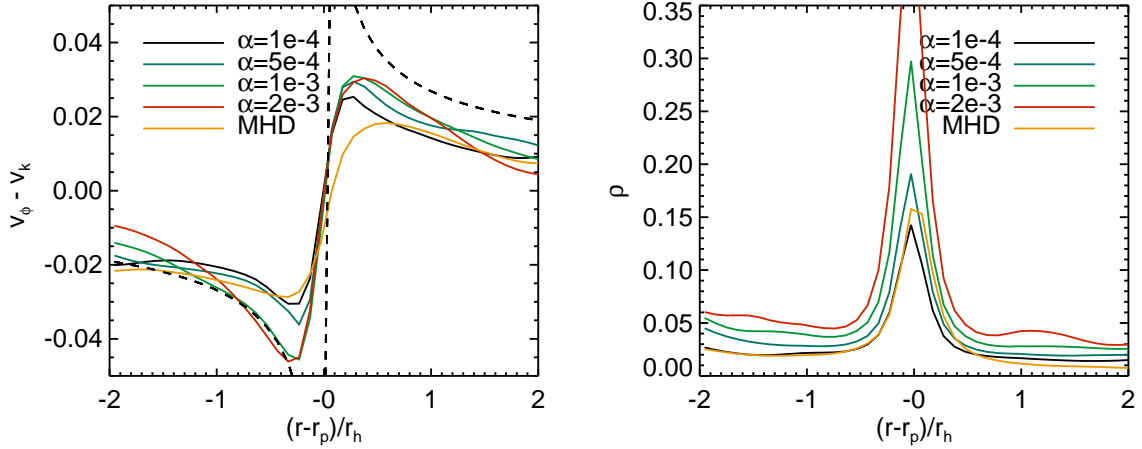}
   \caption{Left: Azimuthal velocity of the circumplanetary gas (relative to Keplerian speed). The dashed line shows the Keplerian speed profile relative to the planet, that has been reduced by 50\%. Right: Density of the circumplanetary gas for the different runs. These figures show radial profiles in a line joining the star and planet. }\label{fig:f2}
\end{figure*}

In this section we study the structure of the density and the inflow of mass in/into the Hill sphere. The density structure of the disk around the planet can be seen in Figure \ref{fig:f1}. The color and contour lines represent the density. The circumplanetary disk is shown for the viscous simulations with $\alpha = 2\times10^{-3}, 1\times10^{-3}, 5\times10^{-4}$, and for the magnetized disk simulation (bottom right panel). The horizontal density contours signal a disk with a scale height increasing linearly with radius from the planet, and with no flaring. Additionally, the density drops with radius away from the planet approximately as $\sim r^{-3/2}$. This is consistent with the density profile found by \citet{2009MNRAS.397..657A} using higher-resolution radiation-hydro simulations. The find the disk profile to be described by a power law between $r^{-2}$ and $r^{-3/2}$. Figure \ref{fig:f2} shows the azimuthal velocity and density of the circumplanetary gas. The velocity profile is consistent with rotation around the planet, and the density profile increases inwards in all cases. We note that the rotation of the circumplanetary gas is significantly sub-Keplerian (with respect to Keplerian motion around the planet, not the star). The azimuthal velocity of the gas in the disk of the planet rotates at less than 50\% Keplerian speed. The velocity structure is also not significantly sensitive to the viscosity, while the density (and therefore the angular momentum) increases with viscosity. The higher density in the large viscosity runs might be due to the higher mass transfer rate from the circumstellar disk to the circumplanetary disk. The larger the viscosity, the larger the amount of mass that the stellar disk can feed to the planet disk, therefore increasing the density surrounding the planet, and possibly the accretion rate of gas onto the planet.
\par
\begin{figure*}[h!]\centering
  \includegraphics[width=\linewidth]{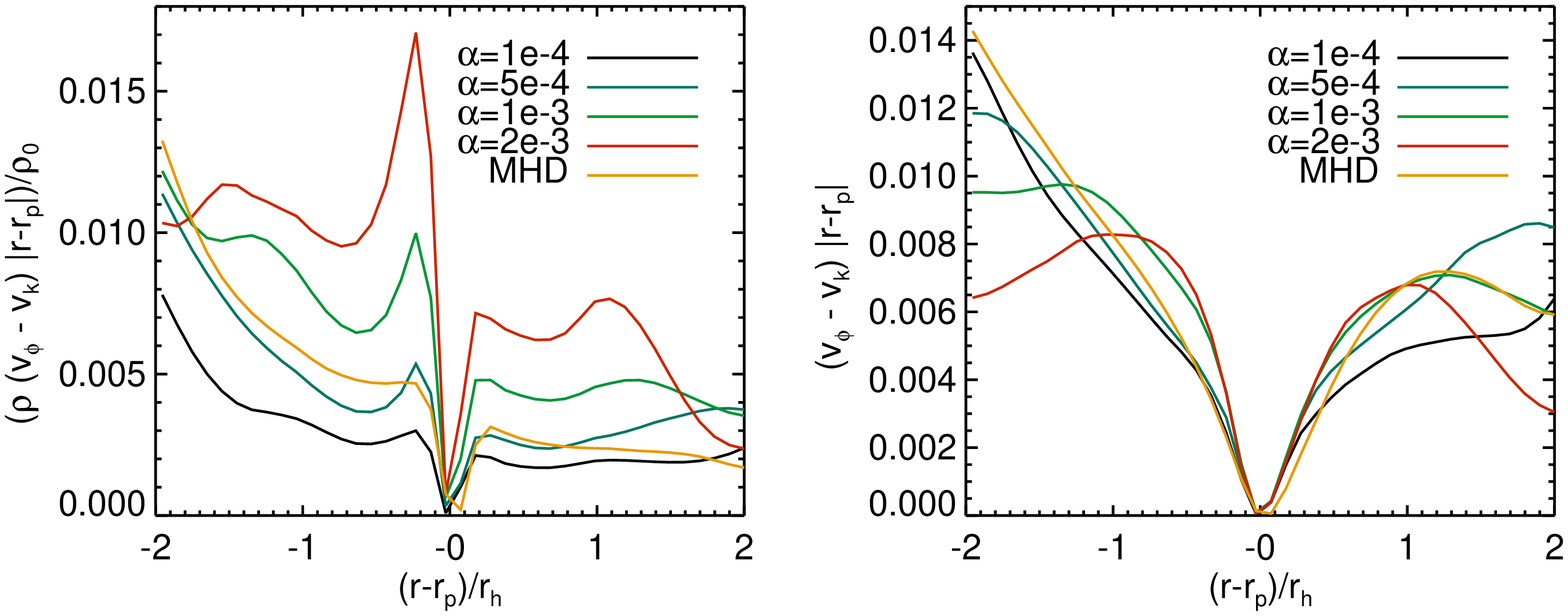}
   \caption{Left: Angular momentum of the circumplanetary gas (normalized to initial densities). Right: Specific angular momentum of the circumplanetary gas for the different runs. These figures show radial profiles in a line joining the star and planet.}\label{fig:f22}
\end{figure*}

The distribution of angular momentum in the circumplanetary disk is shown in Figure \ref{fig:f22}. The left plot of this figure shows the angular momentum scaled with the initial density, while the right plot shows the specific angular momentum. Both quantities are calculated with respect to the planet. The angular momentum distribution increases in the inner part of the disk, and levels off at $\sim 0.2-0.3 r_{h}$. This is consistent with the definition of the edge of the planet disk proposed by \citet{2009MNRAS.397..657A}, since in a Keplerian disk angular momentum increases monotonically outwards (however, at this distances we start to reach the limit of our grid resolution which is $0.1 Hill radii$ in the radial direction). This criterion could define the edge of the Keplerian rotating inner disk, before rotation becomes sub-Keplerian in the outer parts of the disk. However, we also note that we do not obtain Keplerian rotation in this inner part, although this might be affected by a resolution limit. \citet{2009MNRAS.397..657A} obtained a value of $\sim 0.3r_{h}$ for the cutoff radius, and \citet{2012ApJ...747...47T} obtained similar values. This value is also similar to the truncation radius defined by the radius where tidal effects by the central star become important enough to disturb the inner Keplerian rotating disk \citep{2011MNRAS.413.1447M}.
\par
\begin{figure*}[t!]
\begin{minipage}[b]{.5\linewidth}
  \includegraphics[width=\linewidth]{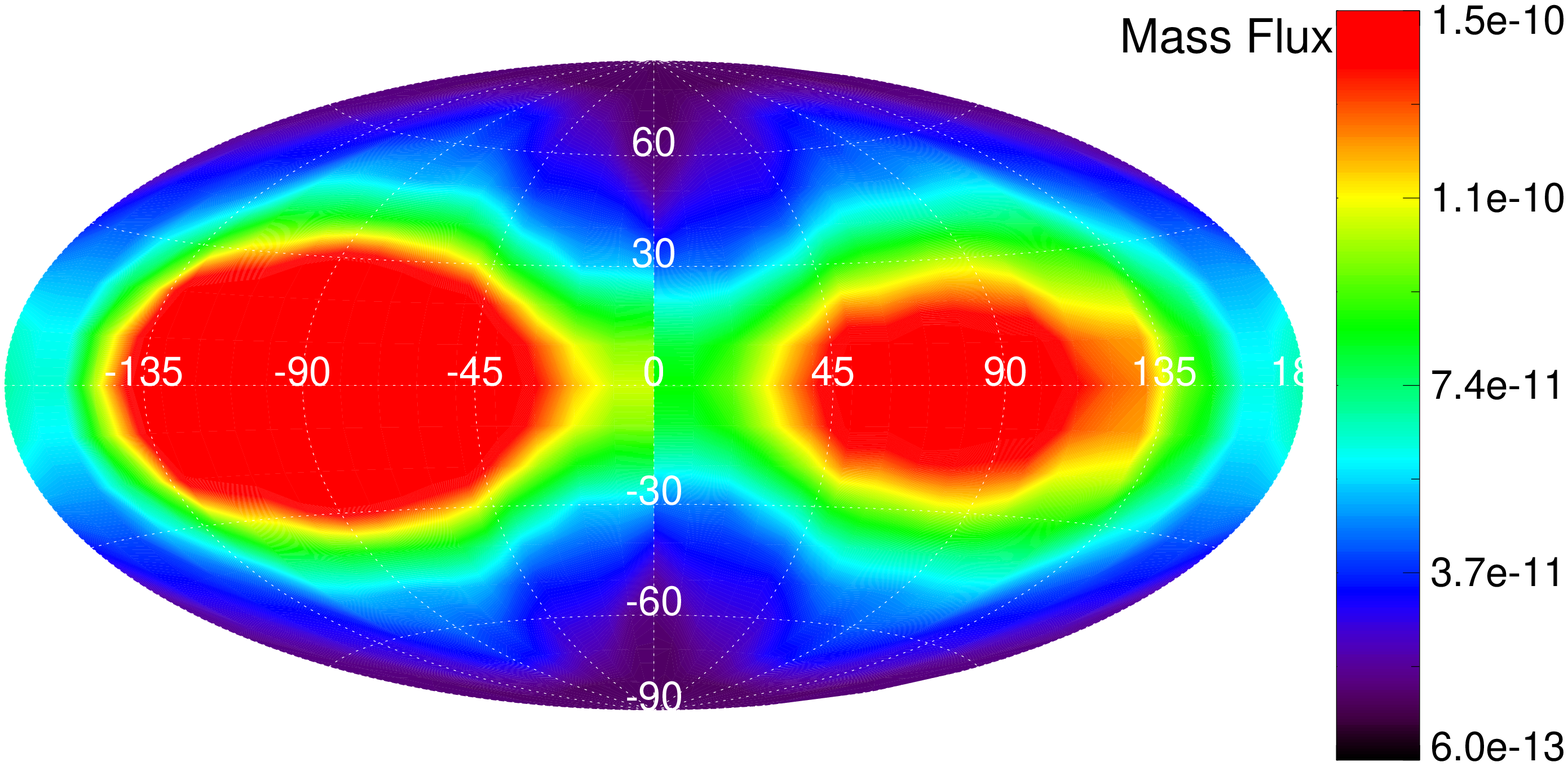}
 \end{minipage}
\begin{minipage}[b]{.5\linewidth}
  \includegraphics[width=\linewidth]{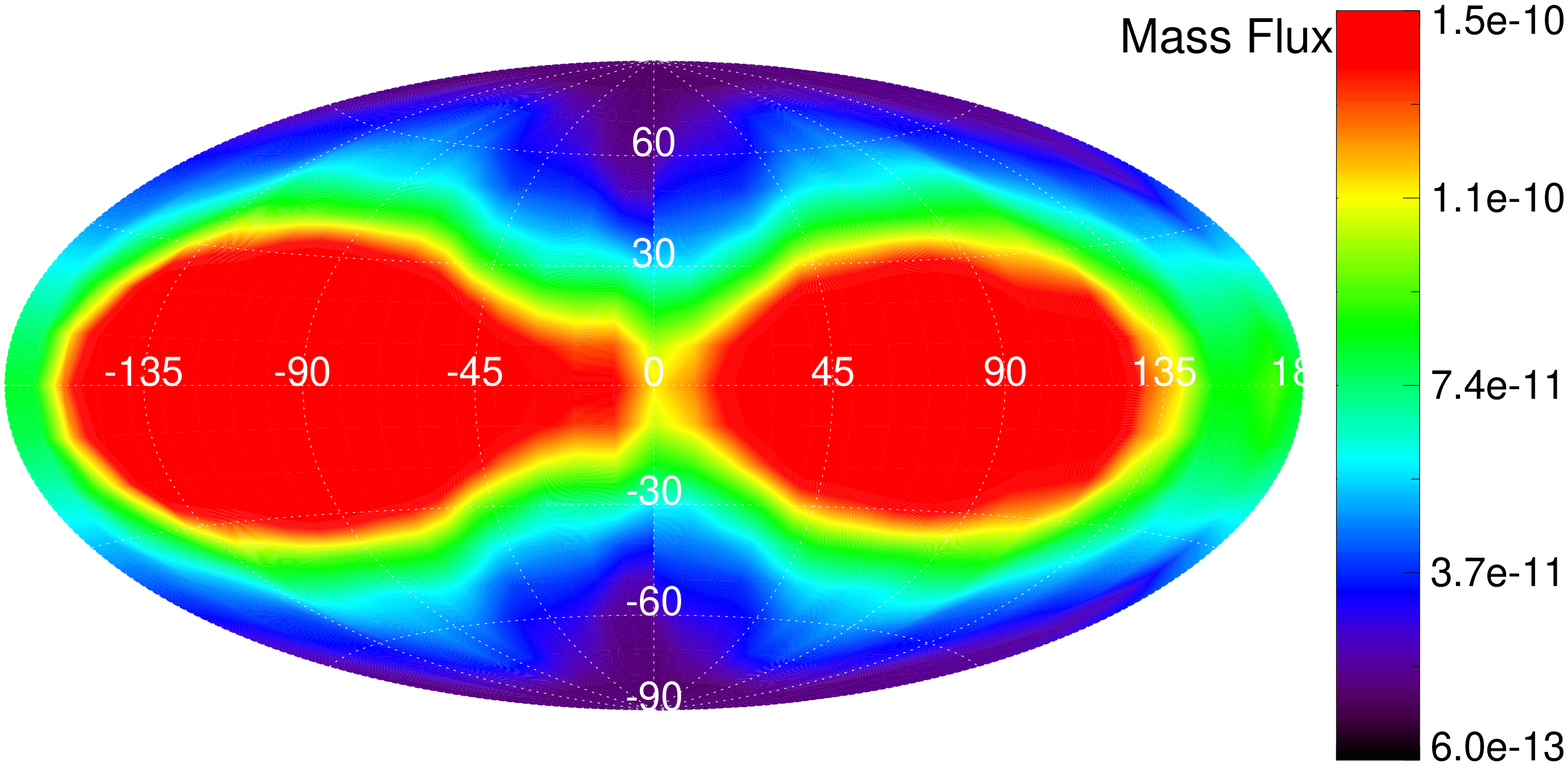}
 \end{minipage}
 \caption{Mass flux through the surface of the Hill sphere for the viscous laminar runs with $\alpha=5\times10^{-4}$(left) and $\alpha=1\times10^{-3}$(right). The mass flux is given in units of $M_{J} yr^{-1} S^{-1}$, where the quantity $S$ is the area of the grid cell in the Hill sphere given by $S=r_{h}^{2}\Delta\theta_{HS}\Delta\phi_{HS}$. The center of the ellipse corresponds to the point in the Hill sphere that is most distant from the star and points away in the radial direction.}\label{fig:f3}

\begin{minipage}[b]{.5\linewidth}
  \includegraphics[width=\linewidth]{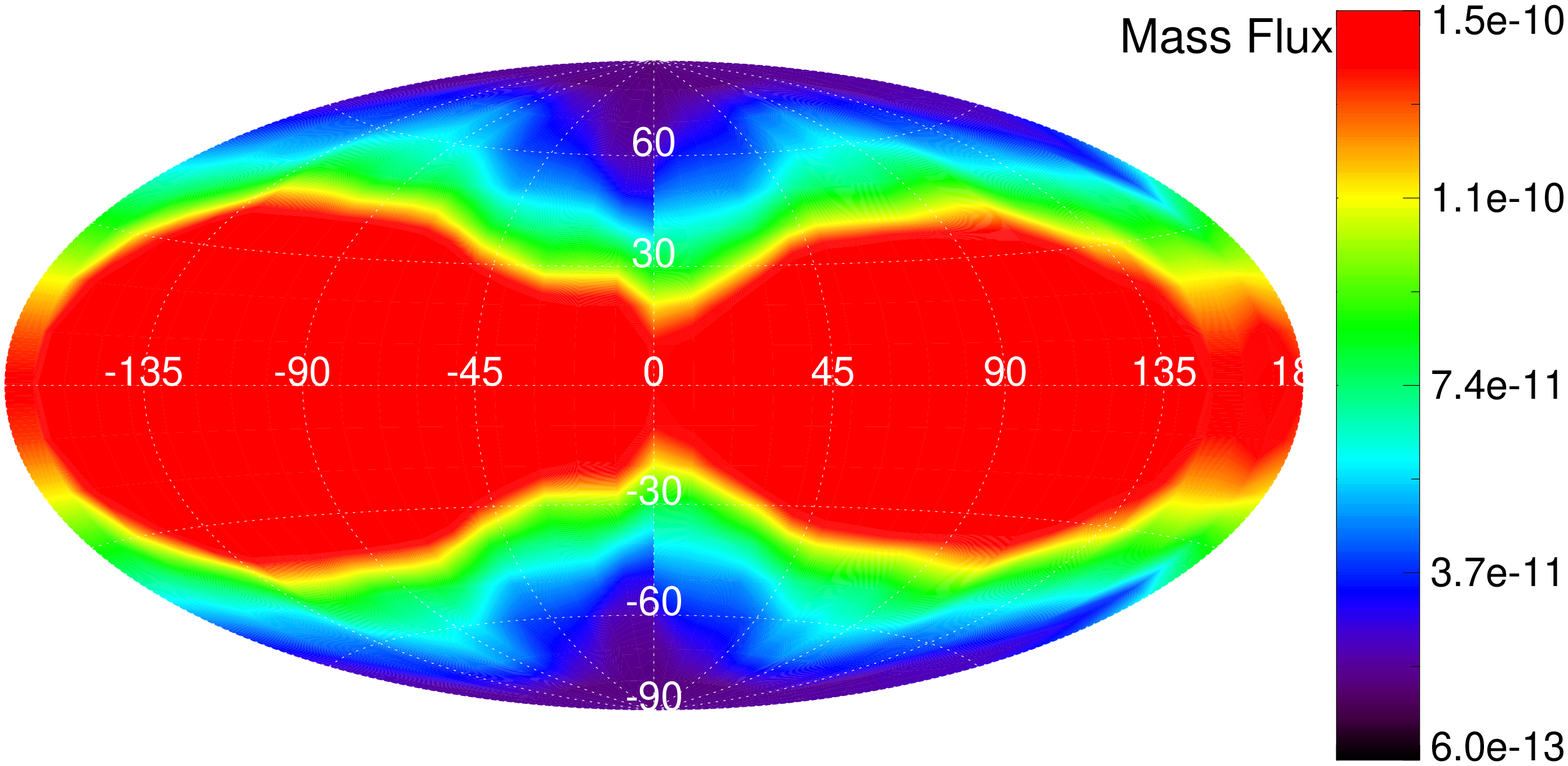}
 \end{minipage}
\begin{minipage}[b]{.5\linewidth}
  \includegraphics[width=\linewidth]{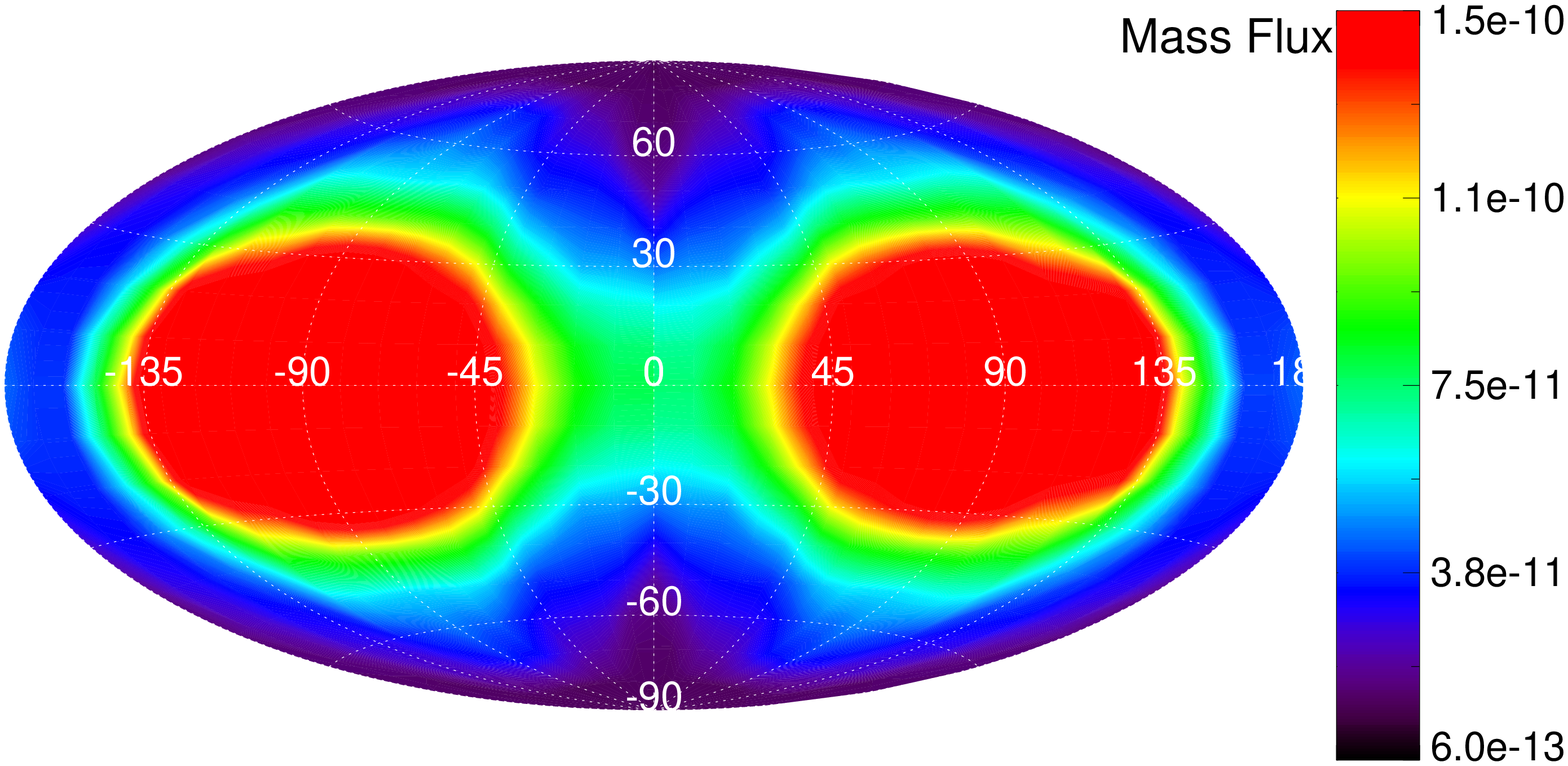}
 \end{minipage}
    \caption{Mass flux through the surface of the Hill sphere for $\alpha=2\times10^{-3}$(left) and for the MHD-turbulent run(right). The mass flux is given in units of $M_{J} yr^{-1} S^{-1}$, where the quantity $S$ is the area of the grid cell in the Hill sphere given by $S=r_{h}^{2}\Delta\theta_{HS}\Delta\phi_{HS}$. The center of the ellipse corresponds to the point in the Hill sphere that is most distant from the star and points away in the radial direction.}\label{fig:f4}
\end{figure*}

 We calculate the mass flux through the surface of the Hill sphere. The mass flux is given by $\rho v_{inflow}=\rho\mathbf{v}\cdot(\nabla F/|\nabla F|)$, where $F(\mathbf{r})=(\mathbf{r}-\mathbf{r_{p}})^{2}-r_{h}^{2}=0$ is the equation describing the surface of the Hill sphere. The mass flux is plotted in Figures \ref{fig:f3} and \ref{fig:f4} for four runs. These quantities have been averaged in time. We plot the surface of the Hill sphere using an ellipsoidal projection where the three-dimensional structure can be observed. The center of the ellipse corresponds to the point in the Hill sphere that is most distant from the star.
\par
\begin{figure*}[h!]\centering
  \includegraphics[width=15cm]{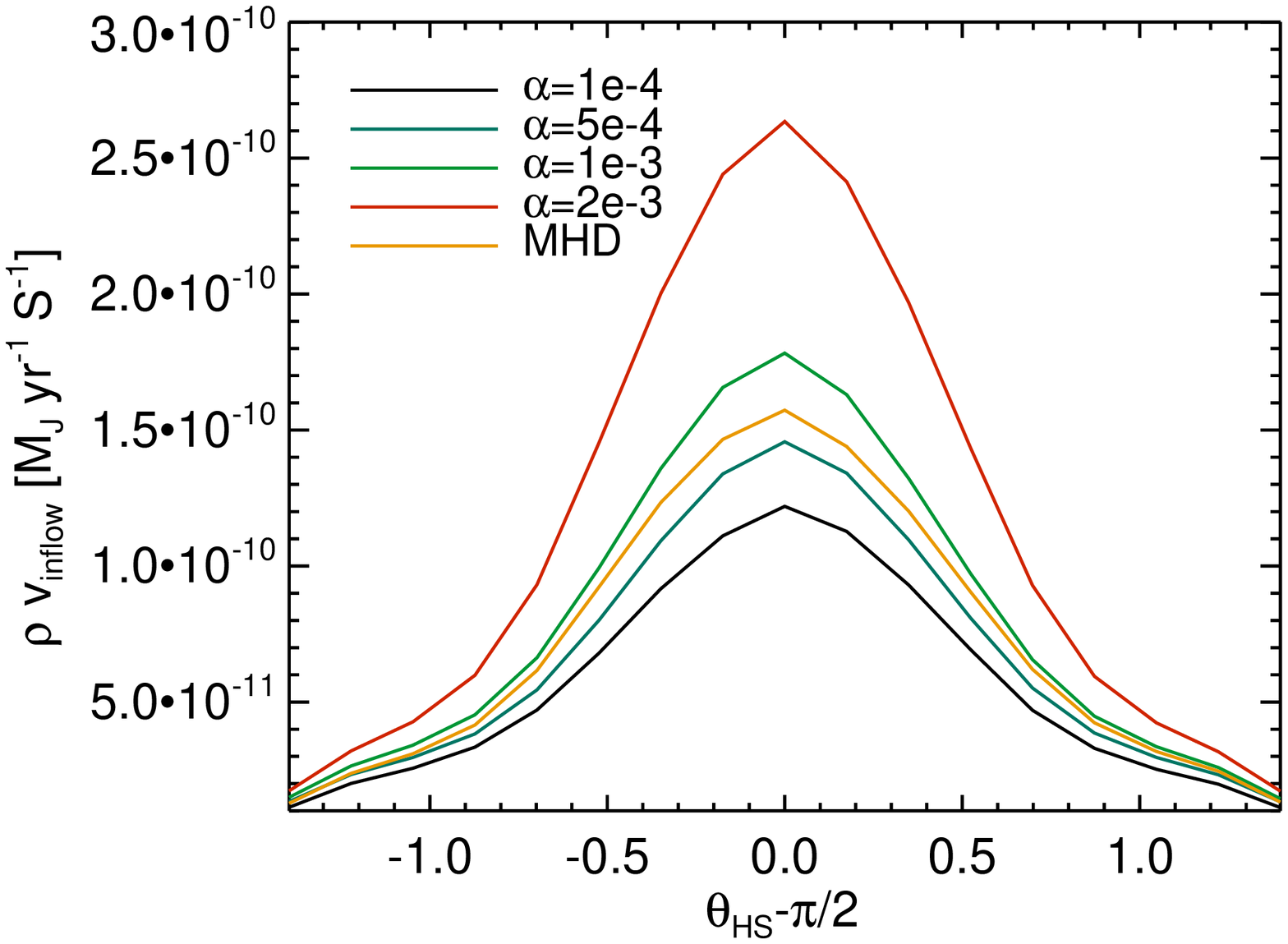}
   \caption{Vertical structure of the mass inflow $\rho v_{inflow}$ into the Hill sphere. The coordinate $\theta_{RH}$ refers to the polar angle in the frame of the planet. The quantity $\rho v_{inflow}$ has been azimuthally averaged (with respect to the Hill sphere). The quantity $S$ is the area of the grid cell given by $S=r_{h}^{2}\Delta\theta_{RH}\Delta\phi_{RH}$. }\label{fig:f44}
\end{figure*}

The density and mass flux are much larger in the case with the higher viscosity ($\alpha=2\times10^{-3}$), and are the lowest in the turbulent magnetized case. In all cases, all the net flux through the surface is inflowing, and we see no significant amount of gas entering the Hill sphere from the vertical direction above and below the mid-plane. The bulk of the accretion at this distance from the planet is confined to the mid-plane "region". Figure \ref{fig:f44} shows the vertical structure of the mass flux averaged over the azimuthal direction. 

\subsection{Mid-plane flow structure in the vicinity of the planet}

\begin{figure*}[h!]\centering
  \includegraphics[width=\linewidth]{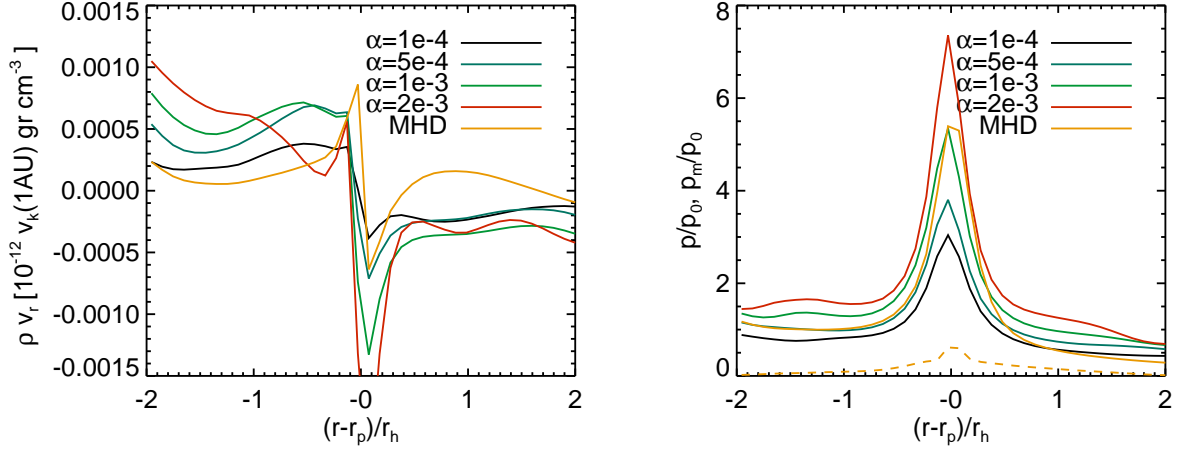}
   \caption{Left: Radial mass flux for the different runs. Right: Pressure for the different runs. The dashed line shows the magnetic pressure (multiplied by a factor of 50) for the magnetized case. }\label{fig:f7}
\end{figure*}

In this section we study the dependence of the flow structure on the viscosity in the simulations and the effect of turbulence on the accretion flow onto the planet. Figure \ref{fig:f7} shows the radial mass flux and the pressure (and magnetic pressure $b^{2}/(8\pi)$) at a distance of $\pm 2r_{h}$ from the planet position. In the mid-plane, the thermal pressure exceeds the magnetic pressure by a factor of the order of $10^{2}$. There is radial inflow of gas coming from both sides of the planet (although inflow is not spherically symmetric in the mid-plane and is not maximum in the line joining the star and planet), and this is dominant in the high viscosity case. As the viscosity gets lower, the density and mass flux decrease. The magnetized case shows radial mass inflow comparable to the two cases with the lower $\alpha$ viscosity, or lower. The radial profile of the thermal pressure is similar for all cases, although the pressure across the Hill sphere decreases with viscosity. At either side of the planet, following the spiral arms, there are bumps of high magnetic pressure. In this region, the magnetic pressure is now comparable to the thermal pressure. 
\par
\begin{figure}[h!]
\begin{minipage}[b]{.5\linewidth}
  \includegraphics[width=\linewidth]{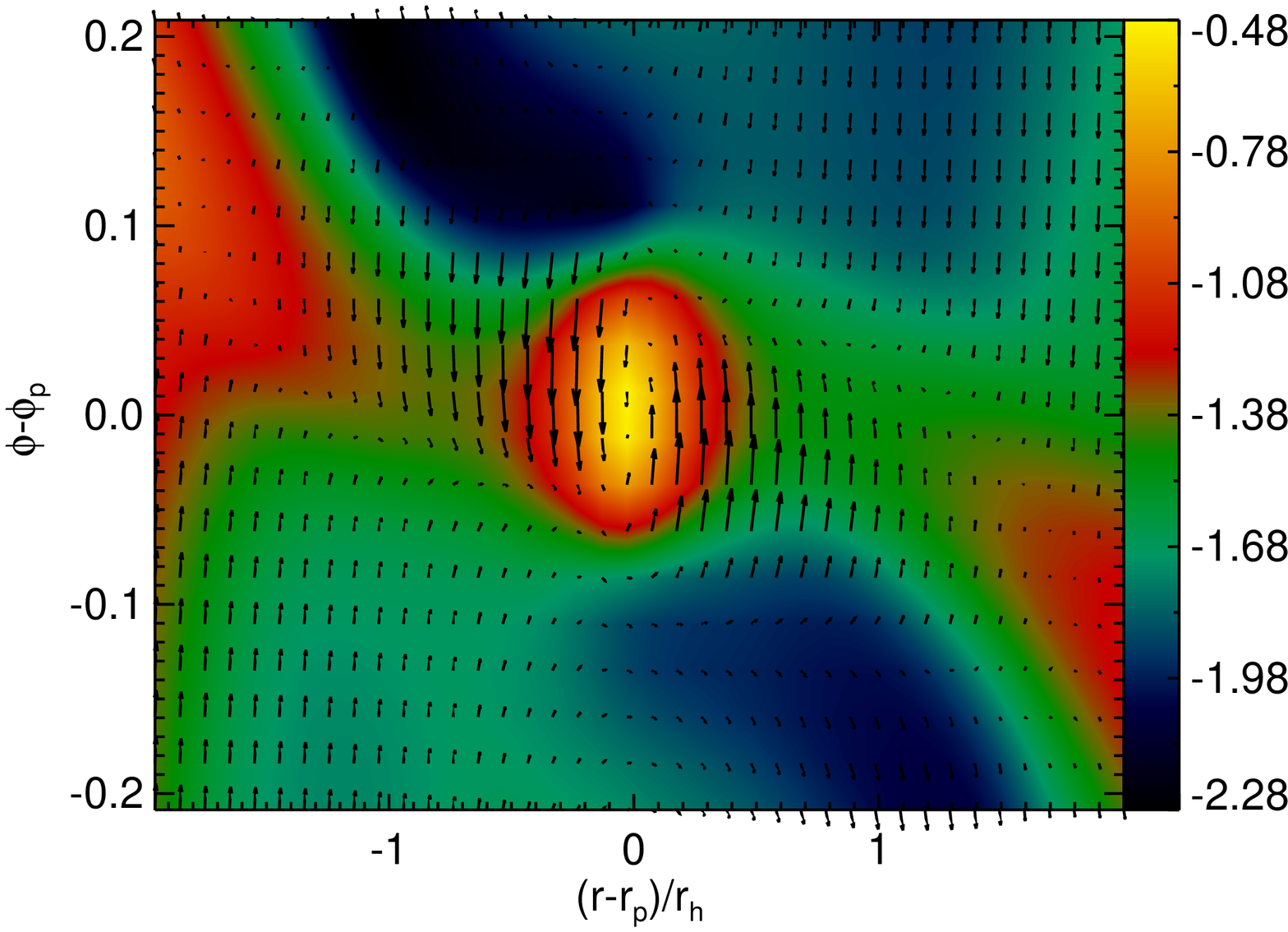}
 \end{minipage}
\begin{minipage}[b]{.5\linewidth}
  \includegraphics[width=\linewidth]{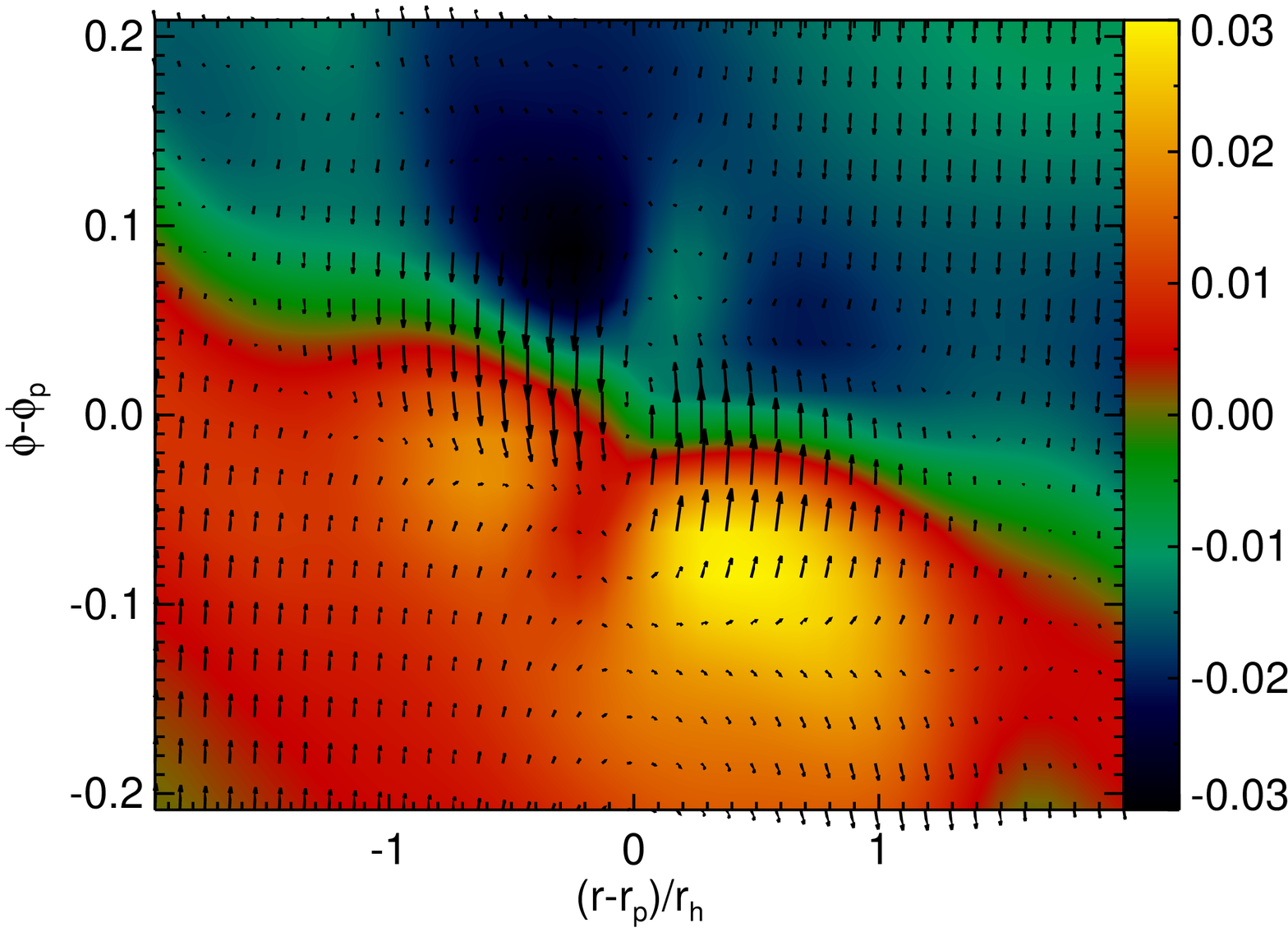}
 \end{minipage}
    \caption{Left:Density (in units of $10^{-12}gr cm^{-3}$) in the mid-plane for the laminar viscous simulation with $\alpha=2\times10^{-3}$. Right: Radial velocity (in units of $v_{k}(1AU)$) in the mid-plane for the same simulation. The overplotted vector field shows the velocity field in the mid-plane.}\label{fig:f8}
\end{figure}
\begin{figure}[h!]
\begin{minipage}[b]{.5\linewidth}
  \includegraphics[width=\linewidth]{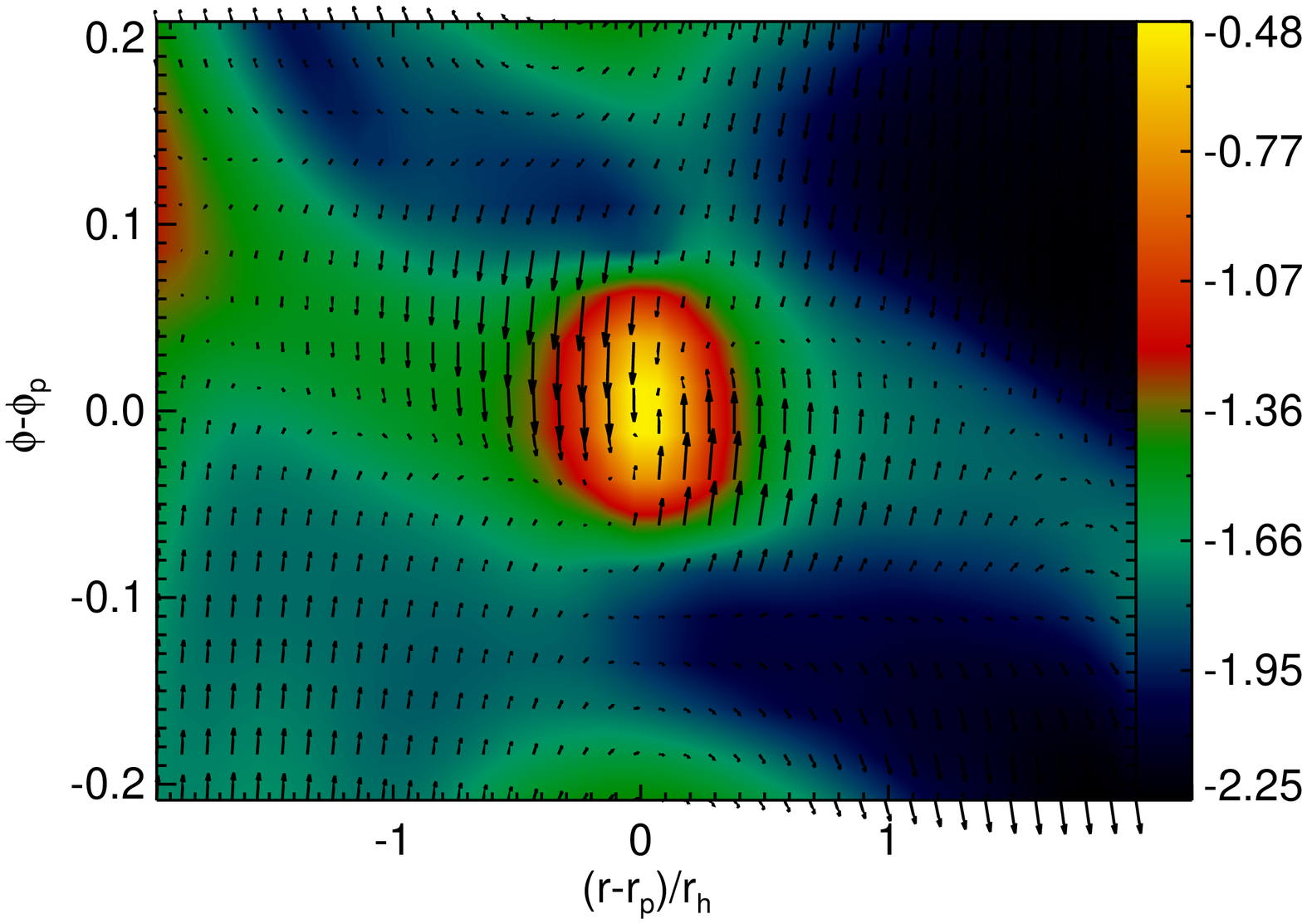}
 \end{minipage}
\begin{minipage}[b]{.5\linewidth}
  \includegraphics[width=\linewidth]{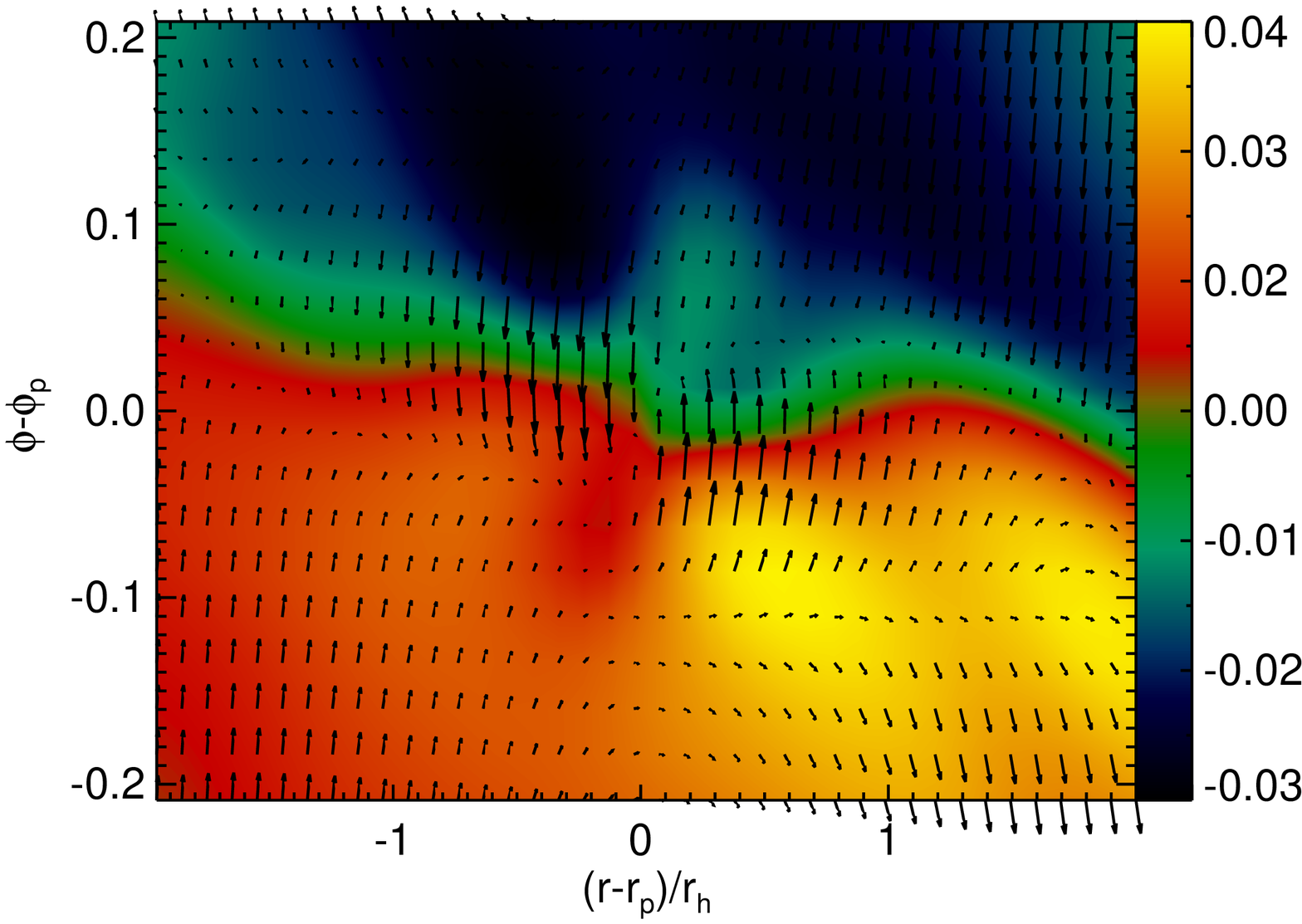}
 \end{minipage}
    \caption{Left:Density (in units of $10^{-12}gr cm^{-3}$) in the mid-plane for the MHD simulation. Right: Radial velocity (in units of $v_{k}(1AU)$) in the mid-plane for the MHD simulation. The overplotted vector field shows the velocity field in the mid-plane.}\label{fig:f9}
\end{figure}

Figure \ref{fig:f8} shows the density in the mid plane and the radial velocity for the laminar viscous simulation with $\alpha=2\times10^{-3}$. Overplotted is the mid-plane vector field of the velocity. Figure \ref{fig:f9} shows the same quantities for the magnetized turbulent circumstellar disk simulation. In agreement with previous studies \citep{2002ApJ...580..506T}, we find that the gas accreting onto the planet comes from a flow between the open pass-by flow and the gas that is orbiting in horseshoe orbits at corotation. This comes from both sides of the planet, as can be seen in the velocity field in the upper right and lower left part of Figures \ref{fig:f8} and \ref{fig:f9}. Material enters the Hill sphere through these channels, as it is also seen in the mass flux at the surface of the Hill sphere in Figures \ref{fig:f3} and \ref{fig:f4}. Inside the Hill sphere, the gas accretes onto the planet through the circumplanetary disk. Two-dimensional inviscid simulations show accretion in the planet disk that is powered by a spiral shock where gas looses energy and spirals inwards. This is not observed in our simulations, since in three-dimensions shocks are smoothed out and the resolution in the disk is not sufficient. Outside the Hill sphere, we see the spiral arm structure that forms the bow shock (see \citet{2002ApJ...580..506T,2003ApJ...586..540D}), although the shock is smeared out by viscosity and turbulence in our simulations. 
\par
In the case of the magnetized turbulent run, the velocity structure around the planet is much less uniform in comparison with the laminar viscous run. This is due to small scale turbulence and the non-uniformity of the magnetic field. However, the mean velocity structure, averaged in time, is similar for both cases as is shown in Figures \ref{fig:f8} and \ref{fig:f9}. 

\subsection{Gas accretion rates and planet growth time after gap-opening}

\begin{figure*}[h!]\centering
  \includegraphics[width=15cm]{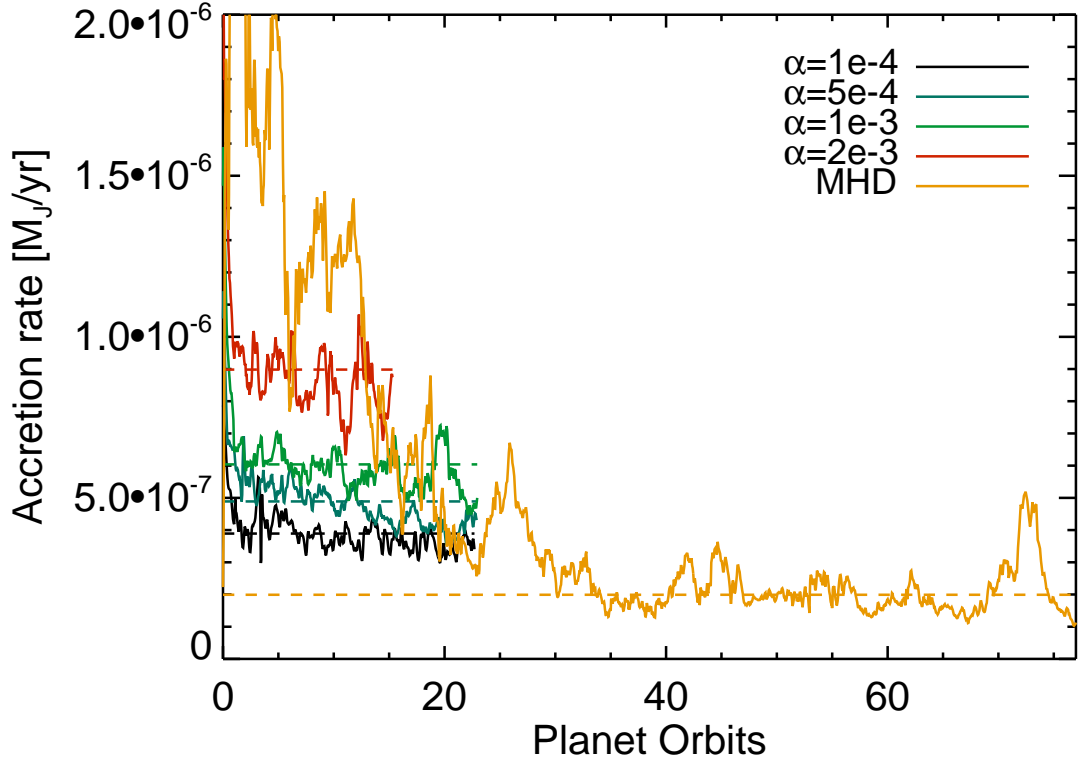}
  \caption{Mass accretion rate onto the planet for the four viscous simulations and the magnetized simulation. The yellow line shows the MHD case. The colored dashed lines show the mean value of each simulation (MHD rates are averaged after 30 orbits).}\label{fig:f5}
\end{figure*}

\begin{figure*}[h!]\centering
  \includegraphics[width=15cm]{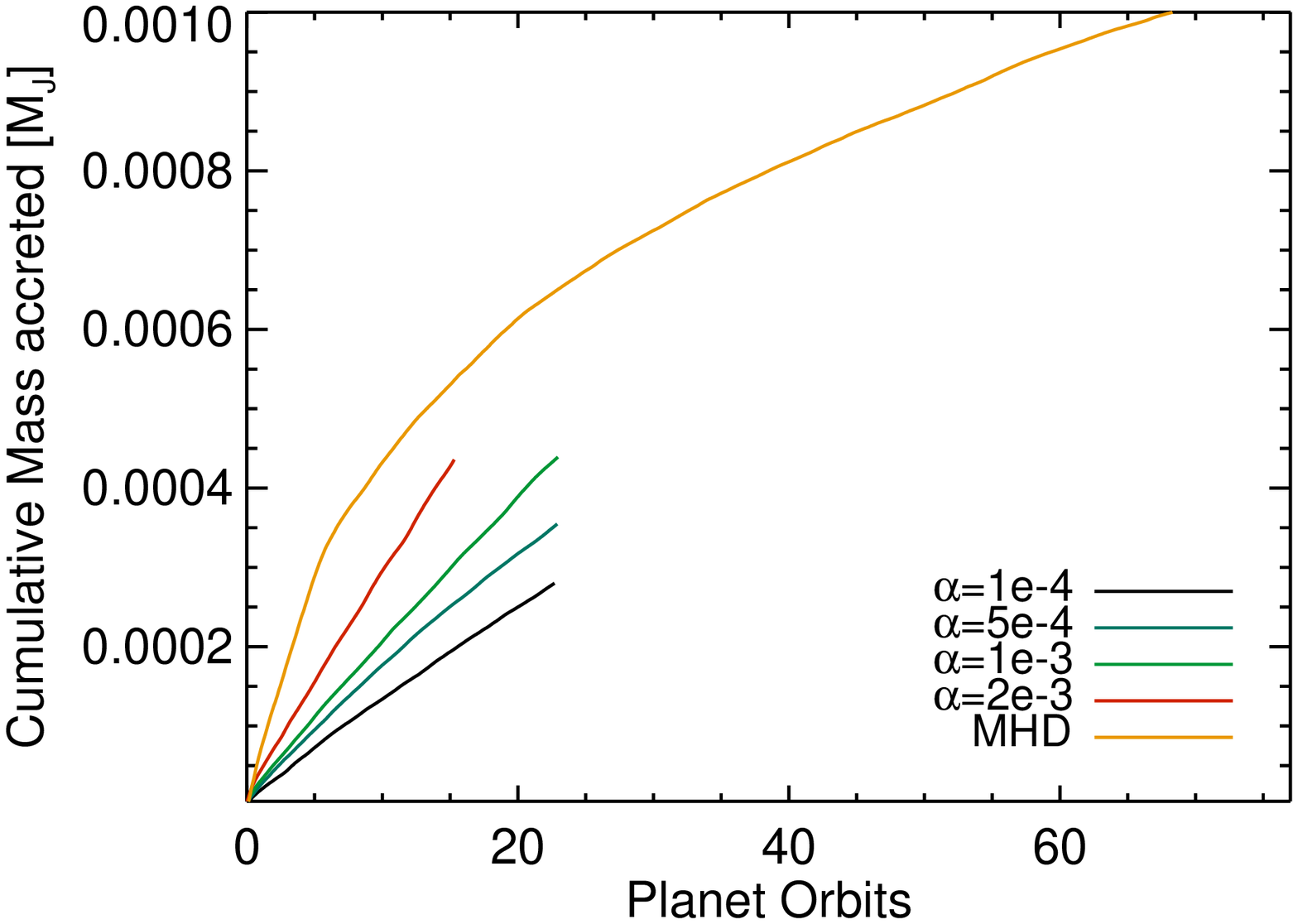}
  \caption{Cumulative mass accreted by the planet for the four viscous simulations and the magnetized simulation. }\label{fig:f6}
\end{figure*}

The cumulative accreted mass and the accretion rate onto the planet are shown in Figures \ref{fig:f6} and \ref{fig:f5}, for the laminar viscous simulations and the magnetized simulation. We will refer to $\alpha_{MHD}$ to denote the alpha stress that is \emph{measured} in the magnetized simulation, and to $\alpha$ to denote the \citet{1973A&A....24..337S} viscosity parameter that is \emph{chosen} for the viscous laminar simulations. The measured alpha is calculated as sum of the contributions to the internal stress by the Reynolds $T_{R}=<\rho\delta v_{\phi}\delta v_{r}>$ and Maxwell magnetic stress $T_{M}=<B_{\phi}B_{r}/(4\pi P)>$, where the brackets denote the mean value over the computational domain (excluding magnetic buffer zones) and $P$ is the pressure. The total stress is additionally normalized by the density. The velocities $\delta v_{r,\phi} = v_{r,\phi} - <v_{r,\phi}>$ denote deviations from the mean value.
\par
We will first discuss the laminar simulations. The largest planet accretion rate is obtained for the viscous simulation with $\alpha=2\times10^{-3}$, as it is expected since the stellar disk accretion rate is proportional to the viscosity in the more simple analytical approximation. The limit of the lowest viscosity that the code is able to resolve above numerical dissipation effects is $\alpha\approx10^{-4}$. Additionally, at this low viscosity, our simulation time is not able to cover the viscous evolution, since the viscous timescale is given by $\tau_{visc}=H^{2}/\nu=H^{2}/(\alpha H^{2}\Omega)=(\alpha \Omega)^{-1}$.
\par
The magnetic case shows an interesting behavior. We find the average planet accretion rate in the magnetic case to decrease drastically during the first stages of the simulation. It seems to stabilize at a value of the accretion rate similar all the viscous simulations, but still presents more variability than the purely viscous cases. In the magnetized case, the planet accretes gas at a rate which is around a third of the accretion rate in an disk with $\alpha=1\times10^{-3}$. For the turbulent magnetized simulation, the global and time-averaged $\alpha_{MHD}$ is equal to $\alpha_{MHD}=1\times10^{-3}$. The global average Maxwell stress is $\alpha_{MHD,Max}=2\times10^{-4}$. Due to the presence of the giant planet, the Reynolds stress dominates over the Maxwell stress by a factor of 2 to 3 \citep{2003ApJ...589..543W,2011ApJ...736...85U}. However, the effective viscosity provided by the turbulence in the mid-plane is less than in the upper layers of the circumstellar disk \citep{2011ApJ...735..122F}. Small scale turbulent structures in the mid-plane might not be well resolved and the $\alpha$ parameter measured for magnetic turbulence measures large scale transport. Nevertheless, the effective viscosity in the mid-plane should be comparable to the one in the viscous laminar simulation with $\alpha=1\times10^{-4}$ \citep{2011ApJ...735..122F}. 

Figure \ref{fig:f6} shows the total mass accreted by the planet starting at a time when the planet accretion is switched on. There is an initial rapid rise due to the material that has been accumulated in the Hill sphere during the previous evolution of 100 orbital periods. After this stage, the planet has consumed the "excess" of material, and accretes at a rate by which the circumplanetary disk can provide material. It can be seen in Figure \ref{fig:f6} that even though the initial conditions are slightly different (some simulations have more gas accumulation around the planet depending on viscosity, as seen in Figure \ref{fig:gap}), after the initial phase is passed, the planet accretion tends to relax to more or less steady state values. 
\par
Previously, \citet{2002ApJ...580..506T} and \citet{1999ApJ...526.1001L} found growth times of $4\times10^{4}$yr. \citet{2001ApJ...547..457K} found gas accretion rates onto the planet of $\dot{M}=6\times10^{-5}M_{J}/yr$ for Jupiter mass planets, which indicates a growth time of $~2\times10^{4}$yr. \citep{2009MNRAS.393...49A} calculated accretion rates for various values of the planet mass, using radiation-hydrodynamics simulations. For a planet with mass equal to the mass we use in our study (a Jupiter mass), they find an accretion rate onto the planet of $\approx 3 \times 10^{-5}M_{Jup}yr^{-1}$.  We find a mass accretion rate of $\dot{M}\approx9 \times10^{-7}M_{J}/yr$ for the laminar viscous case with $\alpha=2\times10^{-3}$. For the magnetized simulation we find $\dot{M}\approx 2\times10^{-7}M_{J}/yr$. These rates are measured only when the planet has already cleared a gap around its orbit, contrary to the previous studies. Our surface density at the position of the planet is therefore very small, as compared to previous studies. We measure a surface density after gap opening of $1.2 g cm^{-2}$ at the position of the planet. Scaling our accretion rates to the surface density value used by \citet{2009MNRAS.393...49A}, the accretion rate we measure is $3\times10^{-5}M_{J}/yr$, which is similar to what was obtained by \citet{2009MNRAS.393...49A}. The scaling of the accretion rate is possible, since the quantity should vary linearly with surface density (in the simple viscous model).

\section{DISCUSSION AND CONCLUSIONS}\label{conc}

\begin{figure}[h!]\centering
  \includegraphics[width=15cm]{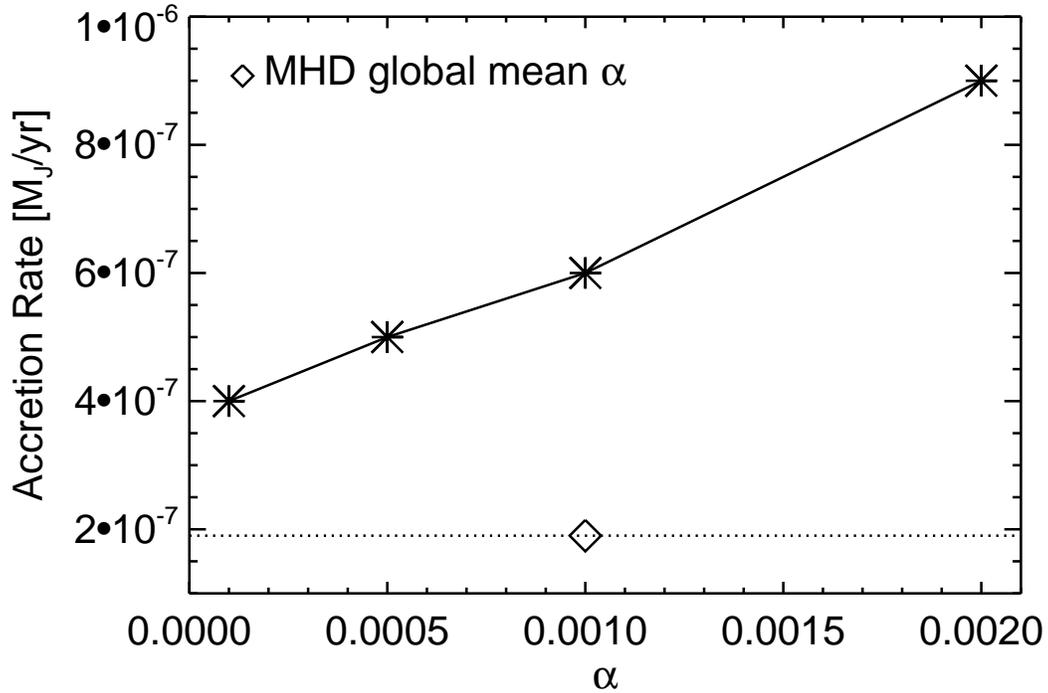}
  \caption{Mass accretion rates by the planet for different values of $\alpha$(crosses) and the turbulent run (diamond and dotted line). The diamond corresponds to the global average measure of $\alpha_{MHD}$ in the magnetized disk simulation. However, this quantity varies vertically and decreases towards the mid-plane.}\label{fig:f11}
\end{figure}

Figure \ref{fig:f11} shows the planet accretion rate as a function of the circumstellar disk $\alpha$ for the laminar viscous runs. The rate obtained in the magnetized run is shown as a dotted line since $\alpha$ is not constant in this case, and the diamond symbol signals the global stress at the beginning of the simulation. For the magnetized case, after 100 orbital periods, the turbulence has decayed as was seen in the simulations presented in \citet{2011ApJ...736...85U}. The effective global averaged stress coming from the turbulence at the beginning of the simulation (right before planet accretion starts) is $\alpha_{MHD}=1\times 10^{-3}$. The Maxwell stress at the beginning of the run is $\alpha_{MHD,Max}=2\times10^{-4}$. In the magnetic case, the measured planet accretion rate is smaller than in all laminar viscous runs and it is a third of the accretion rate in the viscous run with $\alpha \approx 1\times 10^{-3}$. This can be attributed to the fact that turbulent transport is achieved mainly at the large scales, while the effective viscosity provided by the turbulence at the small scales is not represented by the global value of the measured $\alpha_{MHD}$. There also remains the question of how well small-scale turbulence is resolved in our simulations. It is possible that higher resolution allows for smaller structure turbulence and transport on smaller scales.
\par
The mass accretion into the Hill sphere happens along two channels or bands located at each side of the planet \citep{2002ApJ...580..506T}. Closer to the radial location of the planet, material cannot flow in, and instead performs a U-turn, since it looses its angular momentum rapidly as it approaches the planet. Radially away from the planet, the gravitational torque of the planet is not strong enough to pull the material in fast enough, and instead gas orbits passing the planet. The planet accretion flow lies between these two regions. In all cases, we find that the accretion of gas into the Hill sphere is not spherically symmetric, nor azimuthally symmetric. In all cases, the flow through the vertical direction is negligible and the flow is restrained to low latitudes. In our simulations, the Hill radius is approximately equal to the pressure scale height of the circumstellar disk. 
\par
Our results are consistent with results presented by \citet{2009MNRAS.397..657A}, who performed hydrodynamic radiation simulations of the planet-disk system. They show that the inflow of the planet accretion flow is confined to the mid-plane region, and is negligible at polar latitudes. However, recent simulations by \citet{2007ApJ...667..557T} suggest a different picture, where accretion onto the planet mainly occurs through the vertical direction, and there is no inflow along the mid-plane. In fact, \citet{2007ApJ...667..557T} show that there is significant mass outflow from the Hill sphere in the mid-plane, through the Lagrangian points L1 and L2. This flow structure could be a robust feature of inviscid simulations, as it was also observed before by \citet{2006A&A...445..747K}. \citet{2007ApJ...667..557T} speculate that the main difference between these two results is the presence of explicit viscosity, as we have in the simulations presented in this work. In this case, the circumplanetary disk has an inflow of gas and and outflow of angular momentum. However, contrary to circumstellar disks, we find that the outer circumplanetary disk is very sub-Keplerian. This feature of the disk has also been observed in previous studies \citep{2007ApJ...667..557T,2011MNRAS.413.1447M,2009MNRAS.397..657A}. Unfortunately, we cannot yet achieve such high resolution as in their work, to properly resolve the very inner planet disk ($|r-r_{p}|/r_{h} < 0.1$) where rotation might follow a Keplerian profile. This transition from sub-Keplerian to Keplerian rotation can have important implications for the dynamics of solids in circumplanetary disks. In the outer parts, large particles that orbit with Keplerian speed would feel a large head wind which leads to fast inwards migration. This migration could be significantly reduced at the location where the disk becomes Keplerian, creating the possibility for accumulation and growth of solids in this transitional radius. This is a issue that deserves further study.
\par
The inclusion of the curvature, global gradients, gap opening and a locally isothermal equation of state (all which are included in this study) could also play a role in the difference in flow structure as compared to previous studies, since they break the symmetry observed in the work by \citet{2007ApJ...667..557T}.
We have studied the scenario where the stellar disk turbulence indirectly affects the accretion of gas in the circumplanetary disk, but we have not assumed that the circumplanetary disk is MRI-unstable, since the planet disk may or may not be subject to the MRI \citep{2011ApJ...743...53F,2012ApJ...749L..37L}, and might in fact have a different viscosity and accretion mechanism than the circumstellar disk.

\section{SUMMARY}

For the first time, we study the accretion rate across a gap, onto Jupiter-mass planets, in stellar disks that have turbulence present due to the magneto-rotational instability. We compare our results with the planet accretion rates found in laminar viscous disks, and with results from previous studies, where magnetic fields and turbulence are not explicitly present. Contrary to all previous accretion studies, the turbulence in the disk is self-consistently generated by the MRI. We also focus exclusively on the stage when the gap is cleared. This is a critical stage, since is a factor that determines the limiting mass of giant planets. We also study the structure of the outer parts of the circumplanetary disk that forms around the planet and give estimates that can be implemented in planet population synthesis models. We find that the accretion flow into the Hill sphere of the planet is not spherically or azimuthally symmetric, and is predominantly restricted to the mid-plane region of the stellar disk. Even in the turbulent case, we find no significant flow of mass into the Hill sphere coming from high latitudes. Accretion rates are most closely approximated analytically by using the reduced density in the gap region. This means that the gap-opening planet never reaches an accretion rate as high as the one given by the unperturbed density of the circumstellar disk. In a turbulent magnetized stellar disk with initial global stress parameter of $\alpha_{MHD}=1\times 10^{-3}$, we find a third of the accretion rate onto the planet measured in a laminar viscous disk with $\alpha=1\times10^{-3}$. The accretion rate for a Jupiter mass planet in an MRI-turbulent disk, after gap opening, is $\dot{M} \approx 2\times10^{-6} M_{J} /yr$ for a gap surface density of $12 g cm^{-2}$. This means the growth timescale is around a few million years to double the mass of a Jupiter mass planet. The density profile in the outer parts ($|r-r_{p}|/r_{h} > 0.3$) of the circumplanetary disk can be approximated by $\sim r^{-3/2}$ and the rotation of the gas is highly sub-Keplerian, which has important consequences for the dynamics of solids in the disk around the planet and the formation of satellites.

\acknowledgments

The authors acknowledge the computing time on the Bluegene/P supercomputer and the THEO cluster at the Rechenzentrum Garching (RZG) of the Max Planck Society. A. Uribe would like to acknowledge the support of the International Max Planck Research School (IMPRS) of Heidelberg.

\end{document}